\documentclass[superscriptaddress,amssymb,nofootinbib,longbibliography]{revtex4-2}
\usepackage{amsmath}
\usepackage{hyperref}
\hypersetup{colorlinks=true, urlcolor=blue,linkcolor=blue, citecolor=blue}
\usepackage[utf8]{inputenc}
\usepackage{graphicx}
\usepackage{subcaption}
\usepackage[normalem]{ulem}

\begin{document}    
	
	\title{Energy levels for $\mathcal{PT}$-symmetric deformation of the Mathieu equation}
	
	\author{E. Cavalcanti}
	\email[]{erich@cbpf.br}
	\affiliation{Centro Brasileiro de Pesquisas F\'{\i}sicas/MCTI, 22290-180 Rio de Janeiro, RJ, Brazil}
	\author{N.M. Alvarenga}
	\email[]{natalia\_de\_melo@msn.com}
	\affiliation{Instituto de F\'{\i}sica, Universidade do Estado do Rio de Janeiro, 20559-900 Rio de Janeiro, RJ, Brazil}
	\author{F. Reis}
	\affiliation{Departamento de Matemática Aplicada, Universidade Federal do Esp\'{i}rito Santo, 29932-540 Campus S\~ao Mateus, ES, Brazil}
	\author{J.R. Mahon}
	\affiliation{Instituto de F\'{\i}sica, Universidade do Estado do Rio de Janeiro, 20559-900 Rio de Janeiro, RJ, Brazil}
	\author{C.A. Linhares}
	\affiliation{Instituto de F\'{\i}sica, Universidade do Estado do Rio de Janeiro, 20559-900 Rio de Janeiro, RJ, Brazil}
	\author{J.A. Lourenço}
	\affiliation{Departamento de Ci\^encias Naturais, Universidade Federal do Esp\'{i}rito Santo, 29932-540 Campus S\~ao Mateus, ES, Brazil}
	
	
	\begin{abstract}
		We propose a non-Hermitian deformation of the Mathieu equation that preserves $\mathcal{PT}$ symmetry and study its spectrum and the transition from  $\mathcal{PT}$-unbroken to $\mathcal{PT}$-broken phases. We show that our model not only reproduces behaviors expected by the literature but also indicates the existence of a richer structure for the spectrum. We also discuss the influence of the boundary condition and the model parameters in the exceptional line that marks the $\mathcal{PT}$ breaking.
	\end{abstract}
	\maketitle
	
	\section{Introduction}
	\label{sec:introduction}
	
	The study of $\mathcal{PT}$-symmetric systems was originally born of pure formal curiosity. This category of models is defined by the invariance of the Hamiltonian over the successive application of a parity operator ($\mathcal{P}$) and a time-reversal operator ($\mathcal{T}$). A remarkable property of $\mathcal{PT}$-invariant systems is that they can have real energy levels even though the Hamiltonian is non-Hermitian. Not only that, $\mathcal{PT}$-symmetric Hamiltonians are suited to investigate the boundary between closed and open systems. This is indicated by inspection of the spectrum: we can have a $\mathcal{PT}$-symmetric phase (where all the energy levels real, closed scenario) and a $\mathcal{PT}$-broken phase (where some of the energy levels become complex, a transition to an open scenario). Such change is controlled by some \textit{deformation} parameter in the Hamiltonian, which occurs because the $\mathcal{PT}$ operator is non-linear. Therefore, although it commutes with the Hamiltonian, there is no guarantee that they will share the same eigenstates. This richness motivated a plethora of physical applications in optics, superconductivity, microwave cavity, lasers, electronic circuits, chaos, graphene, metamaterials,..., and the subject has experienced an overwhelming growth in the past twenty years (see references in \cite{Bender:2020gbh,Bender:2016epn,Coulais:2020nature,El-Ganainy:2018nature}).
	
	The investigation of $\mathcal{PT}$-symmetric systems started with the study of the spectrum of a deformed anharmonic oscillator~\cite{Bender:1998ke} and the reality of the spectrum was later rigorously proved~\cite{Dorey:2001}. A common behavior of such systems is that, depending on the ``strength" of the deformation, pairs of energy levels join at a so-called exceptional point. This behavior has been observed since the first examples were considered and it has repeatedly appeared in the literature over the years. Around the same time, it seems that Ref.~\cite{Fernandez:1999}, through the use of a non-Hermitian deformation of a periodic potential ($(i\sinh x)^\alpha\cosh^\beta(x)$), was the first to indicate the presence of what the community nowadays call $\mathcal{PT}$ breaking, although that work did not really discuss the emerging field of $\mathcal{PT}$ symmetry. 
	
	There is also a different kind of behavior, in which the energy levels form a ``energy loop". This simply means that there is a lower and an upper bound determining the region where the energy level is real. The appearance of such a loop was first exhibited in Ref.~\cite{Delabaere:1998b} for the strong coupling regime of the anharmonic oscillator with a linear deformation. Later on, now discussing from the perspective of the growing field of $\mathcal{PT}$ symmetry, this loop behavior was reported for periodic potentials in Refs.~\cite{Bender:2010rq,Bender:2012ds,Jones-Smith:2013aqa}, which use a Hamiltonian with a potential term of the type $ig\cos(\theta)$. The behavior is also found for different periodic potentials as one can see in Refs.~\cite{Meisinger:2012va,Bender:2018book}.
	
	In the present work we deal with a periodic potential as given by the Mathieu equation, which has a high number of applications in both engineering and physics, from classical phenomena (inverted pendulum, trap of charged particles, radio frequency quadrupole,...\cite{McLachlan:1947book}) to quantum phenomena (quantum particle moving in a periodic potential, quantum pendulum,...). We intend to explore the consequences of a non-Hermitian, but $\mathcal{PT}$-symmetric, deformation on the Mathieu equation. In the past few years, different approaches were considered in the literature which proposed some kind of deformation \cite{Bender:2005hf,Bender:2007nj,Bender:2010rq,Bender:2012ds,Makris:2008,Midya:2010,Jones:2011,Lin:2011prl,Graefe:2011pra,Moreira:2013,Brandao:2019,Longhi:2010pra,Kerner:2011vx,Ge:2014opt,Longhi:2019prb,Longhi:2019prl,Xu:2019chaos,Longhi:2019prl,Longhi:2019prb}. In the context of $\mathcal{PT}$ symmetry, the first works to discuss a periodic potential were Refs.\cite{Bender:2005hf,Bender:2007nj}, which considered a Mathieu-like potential $\cos(\lambda \phi)+i\varepsilon\sin(\lambda\phi)$, where the strength of the deformation is controlled by the parameter $\varepsilon$. 
	
	Instead of being a mere theoretical curiosity, a deformed Mathieu potential seems to appear in the study of graphene~\cite{Kerner:2011vx} and is directly used in the context of optics ~\cite{Makris:2008,Musslimani:2008prl,Midya:2010,Jones:2011,Lin:2011prl,Ruter:2010nature,Graefe:2011pra,Moreira:2013,Brandao:2019}. Ref.~\cite{Kerner:2011vx} shows that in order to describe charge carriers moving in curved/bent graphene one can derive a generalized Dirac equation. This equation, when considering a monochromatic sinusoidal bending of the graphene, can be recast as a Mathieu equation with a complex driven parameter.
	
	In the context of optical lattices, the electric-field envelope of the beam (paraxial equation of diffraction) satisfies a Schrödinger-like equation \cite{Makris:2008,Midya:2010,Jones:2011} in which the refractive index takes the place of the `potential'. The scenario with complex refractive index $n=n_R+n_I$ -- where the imaginary component $n_I$ represents gain or loss of energy -- provides a natural setting to perform experiments with non-Hermitian physics~\cite{Makris:2008,Ruter:2010nature}. 
	
	\section{Deformed Mathieu Equation}
	\label{sec:deformed_mathieu}
	
	Let us consider the classical Mathieu equation~\cite{Hille:1997book},
	\begin{equation}
	u''(x) + \left(a-2q \cos(2x)\right) u(x) = 0.
	\label{eq:mathieu_classical}
	\end{equation}
	\noindent In the language of quantum mechanics it can be taken as a time-independent Schrödinger equation, where $a$ is related to the energy levels. It is a differential equation associated to the Hamiltonian operator
	\begin{equation}
	\mathcal{H}_0 = p^2 + 2q \cos(2x).
	\label{eq:hamiltonian_classical}
	\end{equation}
	One of the simplest deformations we can introduce is to add a sine contribution to the potential, so that we still have a periodic potential. To keep some degree of generality we can consider both a parameter $\delta$, controlling the strength of the deformation, and a parameter $j\in \mathbb{N}$ which gives the frequency of the new term (where $j=1$ is the simplest scenario):
	\begin{equation}
	\mathcal{H} = p^2 + 2q\left(\cos(2x)+i\delta \sin(2j x)\right).
	\label{eq:hamiltonian_deformed}
	\end{equation}
	\noindent This operator is non-Hermitian but preserves $\mathcal{PT}$ symmetry as can be readily checked\footnote{The parity operator $\mathcal{P}$ acts on positions as $x \rightarrow -x$ and the time-reversal operator $\mathcal{T}$ acts as $i\rightarrow-i$.  That this sign change amounts to time reversal was shown by Wigner a long time ago as a condition to preserve the Heisenberg algebra under $\mathcal{PT}$ transformations~\cite{Bender:2016epn}.}. The differential equation for our deformed version of the Mathieu equation is then
	\begin{equation}
	u''(x) + \left[a-2q  \left(\cos(2x)+i\delta \sin(2j x)\right)\right] u(x) = 0.
	\label{eq:mathieu_deformed}
	\end{equation}
	
	The above equation is related to some developments in the literature. As mentioned in the Introduction, in the context of curved graphene the authors of Ref.~\cite{Kerner:2011vx} obtained an expression similar to our model for the choice $j=1/2$, but in their setting the boundary conditions are (in our notation) chosen at $x=0$ and $x=2\pi$ (see in Sec.~\ref{sec:energy_levels} our choice of boundary conditions). Also, Refs.~\cite{Makris:2008,Midya:2010,Jones:2011,Musslimani:2008prl} considered a similar equation but with a fixed value of $q$. We discuss a bit more about this at the end of Sec.~\ref{sec:pt_transition}.
	
	Let us comment here that for $j=1$ one can choose a transformation of the coordinate parameter into the complex plane ($x = z + i \alpha$, with a fine-tuned choice of $\alpha$) such that the non-Hermitian Hamiltonian turns into a Hermitian one. The existence of a possible transformation relating non-Hermitian and Hermitian Hamiltonians is well-known\footnote{It seems the relationship between Hermitian and non-Hermitian Hamiltonians was studied by Dyson back in the 1950s.} and has already been considered for periodic potentials~\cite{Ciafaloni:1978nb,Fernandez:1999,Bender:2007nj}. However, as far as we know, no physical explanation was yet provided for it and it seems to be useful only from a mathematical perspective\footnote{This was commented back in 2015 by C. M. Bender at a conference at the Israel Institute for Advanced Studies
		it seems that no substantial improvement has been done so far.}. Notice that after the transformation the new variable $z$ does not belong to the physical (and real) coordinate space ($x \in \mathbb{R}$).
	
	We remark that at $\delta=1, j=1$ the potential reduces to $2qe^{i 2x}$. An intuitive picture using classical orbits is provided by Ref.~\cite{Bender:2014}, where it is shown that in this case classical orbits are periodic and open. According to the established knowledge on $\mathcal{PT}$ breaking the energy levels are real if the classical orbits are closed. However, one must notice that the argument was produced for a different scenario. When discussing Mathieu-like equations we are usually restricted to a ``zone" given by the boundary conditions. It then becomes indistinguishable whether the orbit is closed or if it is open and the period has the size of the ``zone". This point of view means that the choice $\delta=1,j=1$ still might produce a meaningful behavior. Indeed, this can be seen for example in Ref.~\cite{Longhi:2010pra}, where a band structure for an exponential potential is discussed. Also, Ref.~\cite{Cannata:1998pla} discusses how such a potential depends on the choice of the contour and produces an example with a set of real eigenvalues.
	
	In what follows we first consider the simpler scenario $j=1$ and discuss how the deformation parameter $\delta$ affects the energy levels and the breaking of $\mathcal{PT}$ symmetry. After this, we discuss the influence of $j\neq1$, and we observe what seems to be a new kind of behavior: the mixing of more than two energy levels.

	\section{Energy levels ($j=1$)}
	\label{sec:energy_levels}
	
	One of the simplest ways to get a feeling of how the eigenfunctions behave is to approach the special case $q=0$ (simple harmonic oscillator), for which periodic solutions exist only if $a=n^2$ (with $n=0,1,2,\dots$). To each eigenvalue we can associate pairs of respectively even and odd solutions $\{\cos(nx),\sin(nx)\}$. These can be $\pi$- or $2\pi$-periodic functions. For general $q$, the pairs of solutions which reduce to these as $q\rightarrow 0$ are written in terms of the Mathieu functions $ce_n(x,q)$ and $se_n(x,q)$. Nonetheless, if $q\neq 0$, each of these eigenfunctions has a distinct eigenvalue. Therefore, there is at most one periodic solution for each eigenvalue. The usual notation labels the eigenvalues associated with the even solutions $ce_n(x,q)$ as $a_n(q)$, with $n=0,1,2,\dots$, and those associated with the odd solutions $se_n(x,q)$, with $n=1,2,\dots$, are labelled by $b_n(q)$.
	
	By definition, the eigenvalues of the classical Mathieu equation, Eq.~\eqref{eq:mathieu_classical}, are determined by the following boundary conditions~\cite{Hille:1997book}:
	\begin{subequations}
		\label{eq:boundary}
		\begin{align}
		u'(0)&=u'\left(\pi\right)=0, \qquad &a_{n}(q); \label{eq:boundary_neumann}\\
		u(0)&=u\left(\pi\right)=0,\qquad &b_{n}(q).  \label{eq:boundary_dirichlet}
		\end{align}
	\end{subequations}
	\noindent That is, while the $a_n$ energy levels are given through the application of Neumann boundary conditions at the endpoints, the $b_n$ energy levels are obtained by Dirichlet ones. We shall employ the same boundary conditions to deal with the deformed version of the Mathieu equation, Eq.~\eqref{eq:mathieu_deformed}.
	
	In the scenario without deformation the usual Mathieu spectrum is formed by the curves $a_n(q)$ and $b_n(q)$, separating regions of stability (regions centered in the origin) and regions of instability.
	As soon as the deformation parameter $\delta$ is turned on, what we observe is that we lose a small part of the spectrum, the eigenvalues cease to be real for sufficiently negative values of $q$ (for $\delta=0.1$, see Fig.~\ref{fig:energylevels}, left). 
	As we increase the value of the deformation the allowed region diminishes (for $\delta=0.5$, see Fig.~\ref{fig:energylevels}, middle) and, at some point, we start to see closed energy loops (for $\delta=2$, see Fig.~\ref{fig:energylevels}, right). The contour of these energy loops are the real eigenvalues of the problem and inside this region, we expect stable solutions. 
	
\begin{figure}[h]
		\centering
		\includegraphics[width=0.6\linewidth]{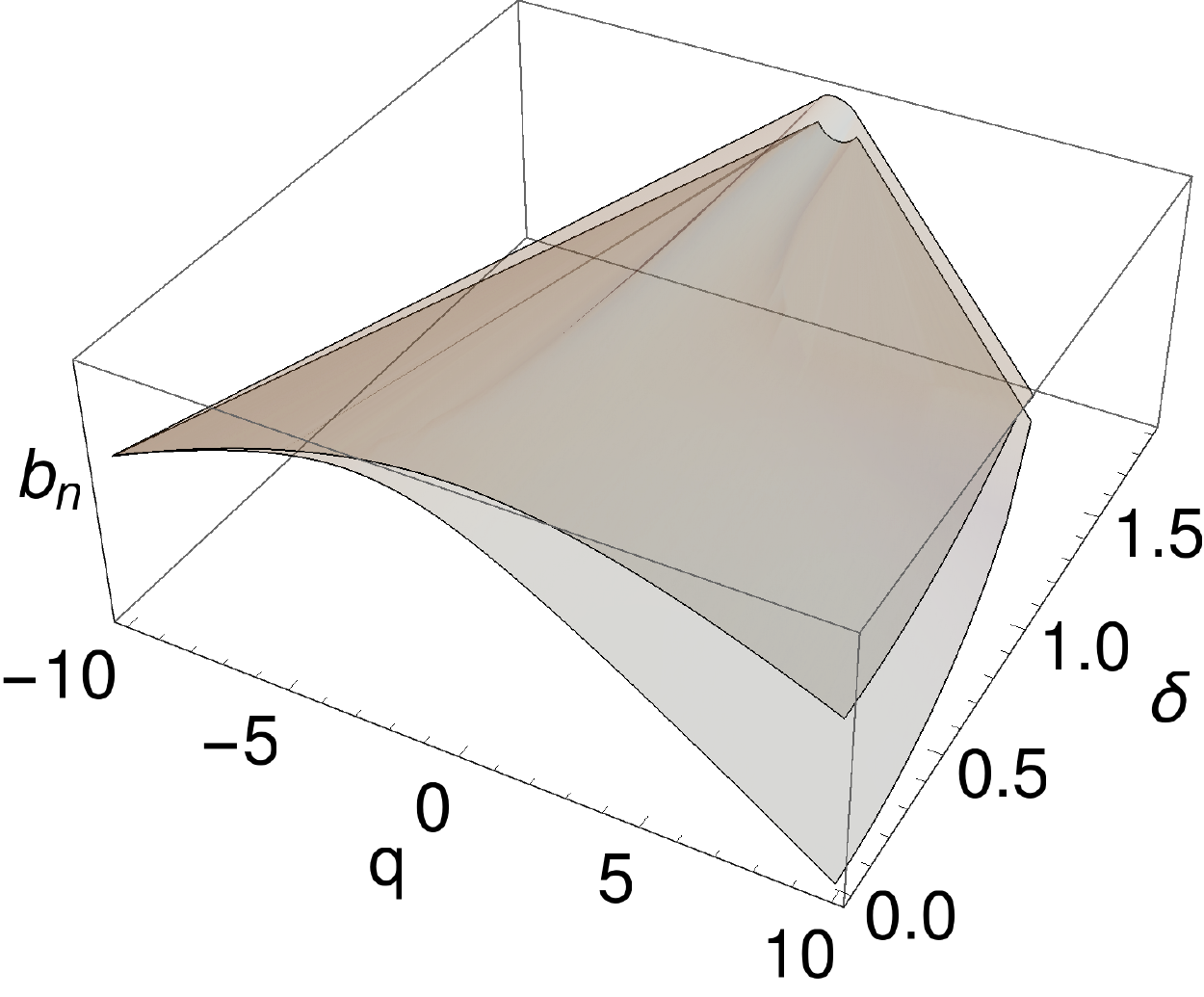}
		\caption{First two energy levels $b_1,b_2$ from the $b_n$ set as a function of both the $q$ parameter and the deformation $\delta$. We can see clearly the behavior from the classical Mathieu (at $\delta=0$) and that the curves seem to smoothly fold as the deformation parameter increases.}
		\label{fig:energylevels_3d}
	\end{figure}

	We can also look directly at the eigenvalues as a function of both $q$ and $\delta$ by a 3D plot, see Fig.~\ref{fig:energylevels_3d} so that we can easily see the smooth transition from the usual Mathieu scenario to the deformed scenario where the allowed energy levels describe loops enveloping a region of stability. 
	
	In Fig.~\ref{fig:energylevels} the smallest energy loop is from the smallest energy level. This behavior should already be expected as in the classical Mathieu those regions of stability \textit{increase} in size when we increase the eigenvalue.
	
	\begin{widetext}
		\quad
		\begin{figure}[h]
			\centering
			\includegraphics[height=4.1cm]{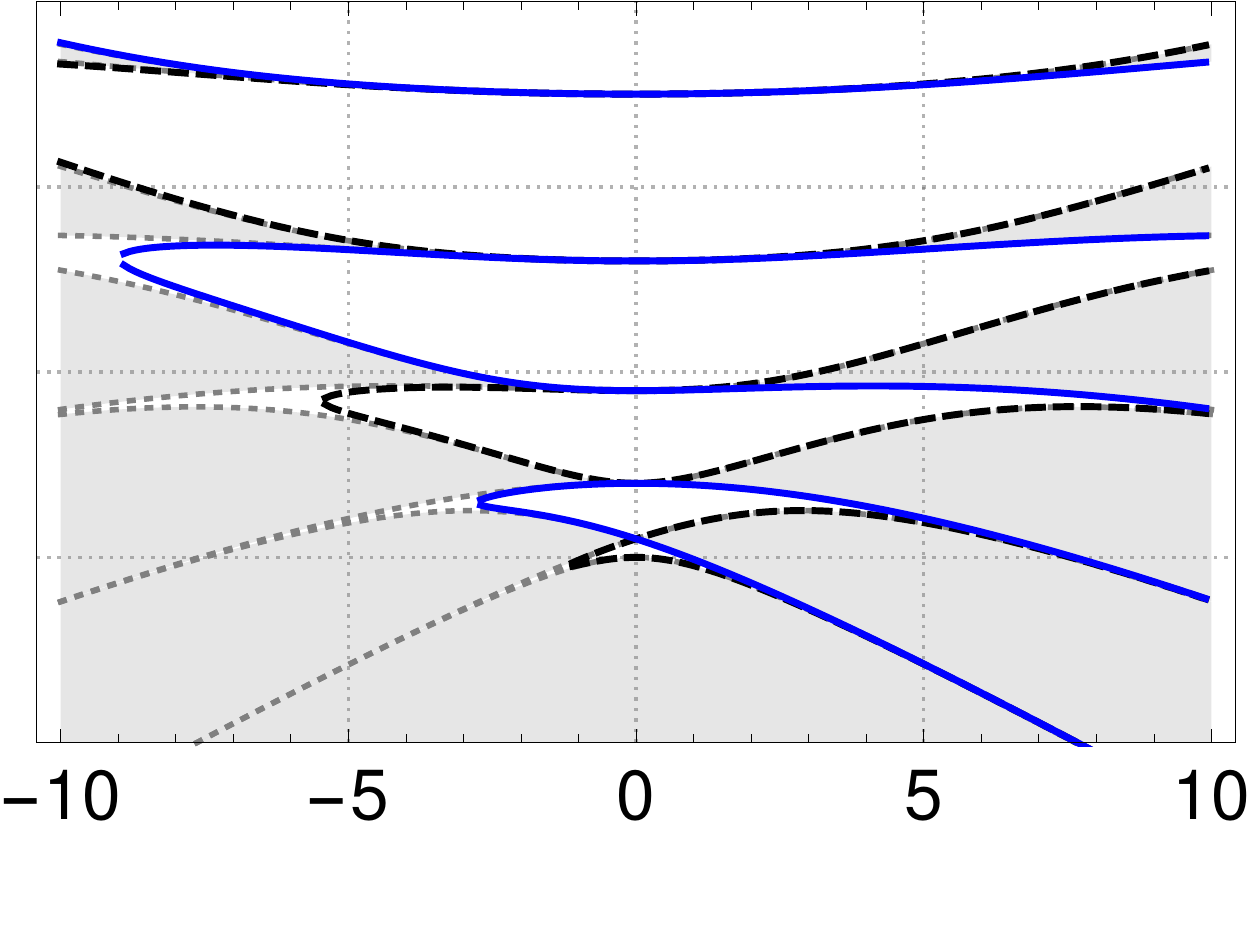}
			\includegraphics[height=4.1cm]{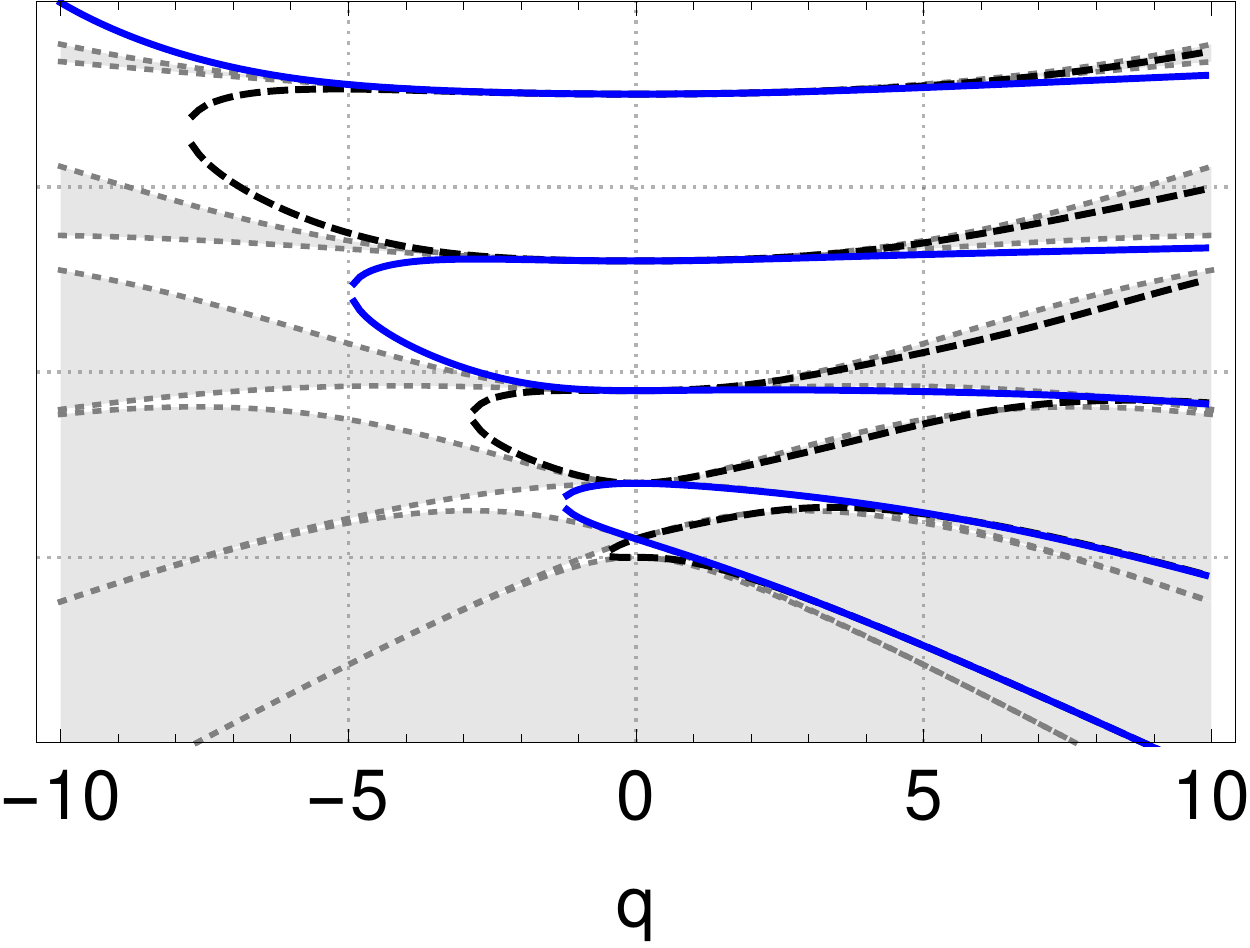}
			\includegraphics[height=4.1cm]{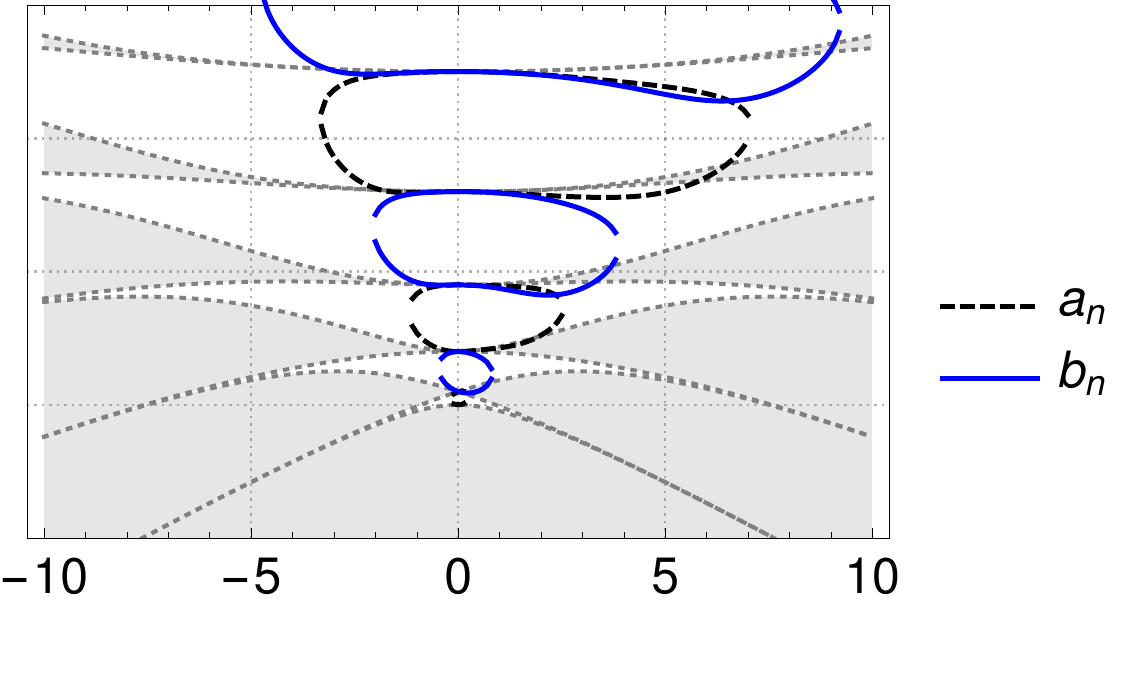}
			\caption{Energy levels for the deformed Mathieu equation. From left to right: $\delta=0.1,0.5,2.0$. The dotted curves are the $a_n,b_n$ curves from classical Mathieu, the gray regions are regions of instability.}
			\label{fig:energylevels}
		\end{figure}
	\end{widetext}
	
	Notice that we lose reflection symmetry in the parameter $q$, in comparison with the classical Mathieu scenario, but there is an invariance under a reflection of the deformation parameter, that is, $a_n(q,\delta) = a_n(q,-\delta)$ and $b_n(q,\delta) = b_n(q,-\delta)$. 
	
	With respect to the loss of reflection symmetry in $q$, one must recall that this relationship was indeed between different sets of energy levels, that is $a_{2k}(q)=a_{2k}(-q)$, $b_{2k}(q)=b_{2k}(-q)$, $a_{2k+1}(q)=b_{2k+1}(-q)$. However, once we turn on the deformation parameter $\delta$ there is a mixing between neighbor energy levels. This is best illustrated in Fig.~\ref{fig:energylevels_3d}, as we can see the evolution of the levels $b_1$ and $b_2$. Already in the undeformed scenario, they seem to asymptotically merge, then when we turn on the deformation the point where they merge (the exceptional point) approaches smaller values of $q$. In other words, the reflection symmetry in $q$ was not really expected in our scenario as it would mean the merging of two independent sets of energy levels ($a_n$ produced by Neumann b.c. and $b_n$ produced by Dirichlet b.c.). 
	
	On the other hand, the reflection symmetry on the deformation seems to indicate an overall balance of the system (at least its spectrum) between gain and loss. If we look back to Eq.~\eqref{eq:hamiltonian_deformed} and the boundary conditions Eq.~\eqref{eq:boundary} we see that $\delta>0$ means that in the first-half region ($x\in[0,\pi/2]$) the hamiltonian has a \textit{positive} imaginary component and in the second-half region ($x\in[\pi/2,\pi]$) the Hamiltonian has a \textit{negative} imaginary component. That is, for $\delta>0$ our system first experiences gain and afterwards experiences loss. When we take $\delta<0$ we change the ordering to loss$\rightarrow$gain. This reasoning suggests that the spectrum of the system does not distinguish whether it first experiences gain or loss.

	\section{PT transition ($j=1$)}
	\label{sec:pt_transition}
	
\begin{figure}
		\centering
		\includegraphics[width=0.6\linewidth]{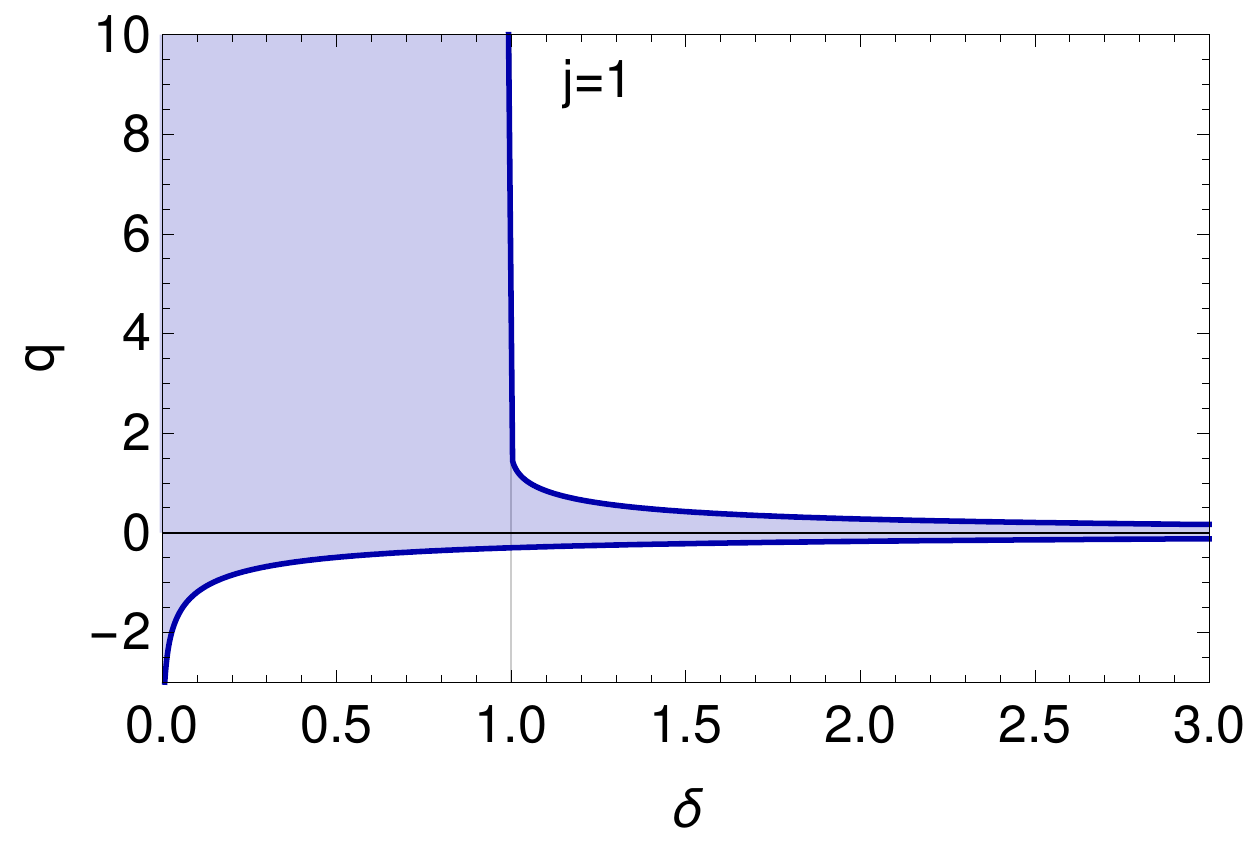}
		\caption{Numerical evaluation of the exceptional line. In the filled region the $\mathcal{PT}$ symmetry remains unbroken and all energy levels are real. Outside this region at least one energy level is lost and $\mathcal{PT}$ symmetry is broken. The line of phase separation is the exceptional line, where a coalescence of energy levels occur. This image has reflection symmetry for $\delta \rightarrow -\delta$.}
		\label{fig:pttransition}
	\end{figure}

	Regarding the reality of the spectrum it is usual to define that the parametric region where the eigenvalues are real is $\mathcal{PT}$ symmetric and the parametric region where at least one eigenvalue is lost is $\mathcal{PT}$ broken. Therefore, the $\mathcal{PT}$ transition occurs exactly when one of the eigenvalues is lost. Our scenario is a bit richer due to the existence of the $q$ parameter. We can, therefore, choose to discuss the $\mathcal{PT}$ transition as the point where \textit{for each value of $q$} at least one eigenvalue is lost. Recalling that the $\mathcal{PT}$ transition means a transition from a closed to an open system, here we are describing a transition from a scenario where the ``ground state" stability is attainable to a scenario where it is completely unstable. This might be tracked by looking at the first eigenvalue, which has the smallest range of stability. So, for each possible value of the deformation parameter $\delta$ we check the value of $q$ ($q>0$) at which the eigenvalues become complex. This can be done numerically and is exhibited in Fig.~\ref{fig:pttransition}. We observe that there is a constraint in the allowed value of the deformation $\delta$, above which the system becomes $\mathcal{PT}$ broken. This upper bound is $\delta=1$ for almost all values of $q$, which seems to be related to the condition we found earlier when we compared the deformed and classical Mathieu equations. Only when this parameter is small, large values of $\delta$ are allowed. This seems to be related to some ``resistance" of the system as it comes near the $q\rightarrow0$ limit, which is related to the harmonic oscillator limit. We refer to ``resistance" in the sense that the system resists undergoing the $\mathcal{PT}$ breaking.
	
	In Fig.~\ref{fig:pttransition} we completely exhibit the phase diagram for $j=1$ according to variations of the parameters $\delta$ and $q$. There are two phases, the unbroken $\mathcal{PT}$ phase being represented by the filled region inside the exceptional lines, and the broken $\mathcal{PT}$ phase being outside the exceptional lines. 
	For negative values of $\delta$ we just apply the reflection symmetry $\delta\rightarrow -\delta$. It is noticeable that the behavior for positive and negative values of $q$ is distinct. While for $q<0$ the phase is already broken for small values of $\delta$, for $q>0$ we stay in the unbroken phase as long as $\delta<1$.
	
	We remark that if someone is interested in a system in which the parameter $q$ varies freely in the positive real axis, it makes sense to say that $\delta=1$ is an upper bound above which $\mathcal{PT}$ symmetry is broken. We can avoid this `restriction' only for small values of the parameter $q$, where larger values of $\delta$ are still allowed.
	
	In the literature, Refs.~\cite{Makris:2008,Musslimani:2008prl,Midya:2010} discuss a potential $4(\cos^2(x)+iV_0\sin(2x))$, while Ref.~\cite{Jones:2011} rewrites it as  $4(\cos(2x)+2iV_0\sin(2x))$. According to them, at $V_0=1/2$ the $\mathcal{PT}$ breaking occurs, but they consider a fixed value of the potential amplitude. If we compare these results with our scenario, we see that it still holds for a choice of $q=2$ for our parameter (this reproduces the fixed amplitude from Refs.~\cite{Makris:2008,Musslimani:2008prl,Midya:2010,Jones:2011}). However, we also generalize it for different values of $q$.
	
	Based on the discussion in the literature~\cite{Makris:2008,Musslimani:2008prl,Midya:2010,Jones:2011} and the findings of the present paper, we suggest an experimental measurement. We propose to consider the effect of the deformation parameter $\delta$ on the $\mathcal{PT}$ breaking when the amplitude $2q$ is reduced. As indicated in Fig.~\ref{fig:pttransition} we expect that the smaller the amplitude the higher the ``stability" of the system. One can understand it intuitively by noticing that the limit of small values of $q$ is equivalent to reducing the system to a harmonic oscillator, which is stable. Therefore, this setting should be more ``resistant" against $\mathcal{PT}$ breaking.
	
	\section{Different scenarios  ($j\neq1$)}
	
	\begin{figure}
		\centering
		\includegraphics[width=0.45\linewidth]{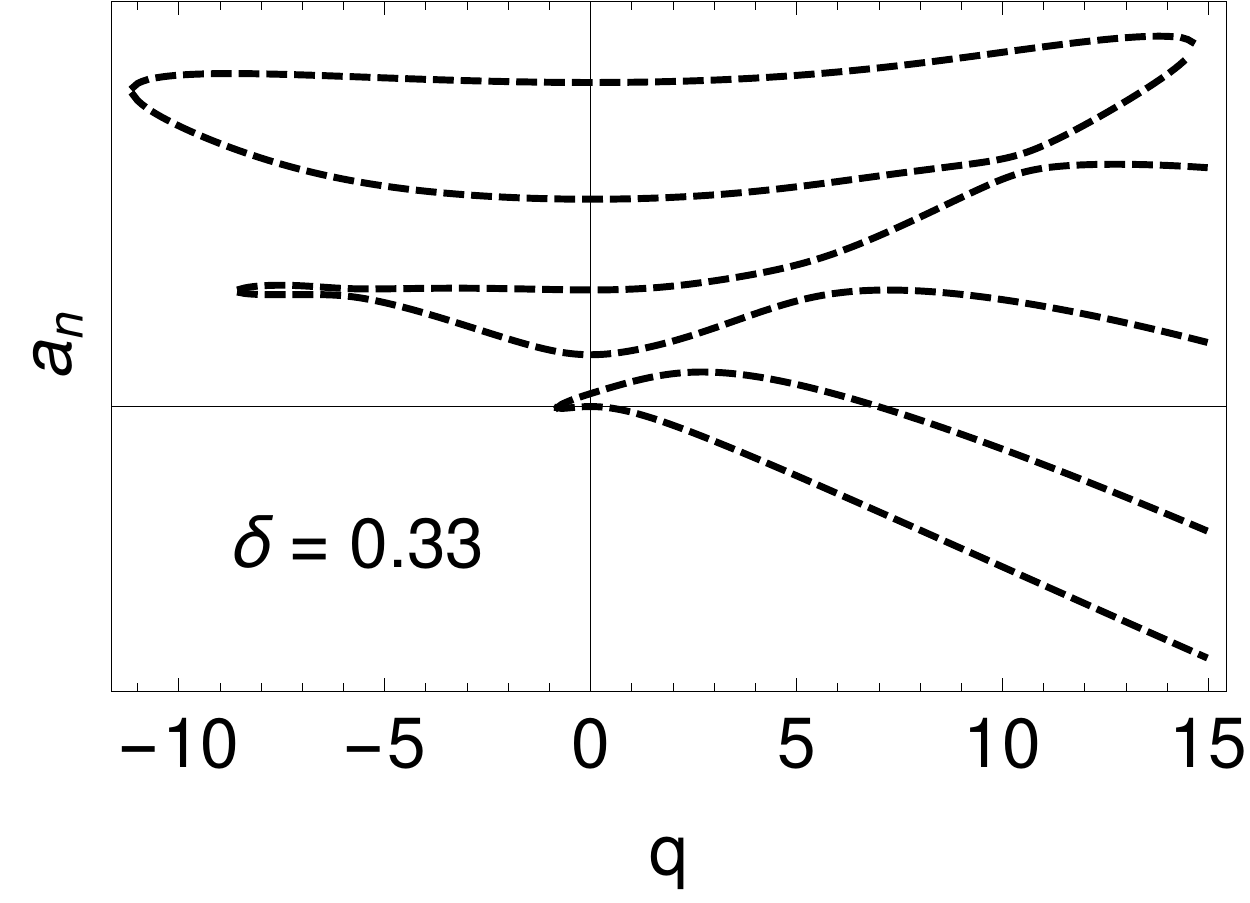}
		\includegraphics[width=0.45\linewidth]{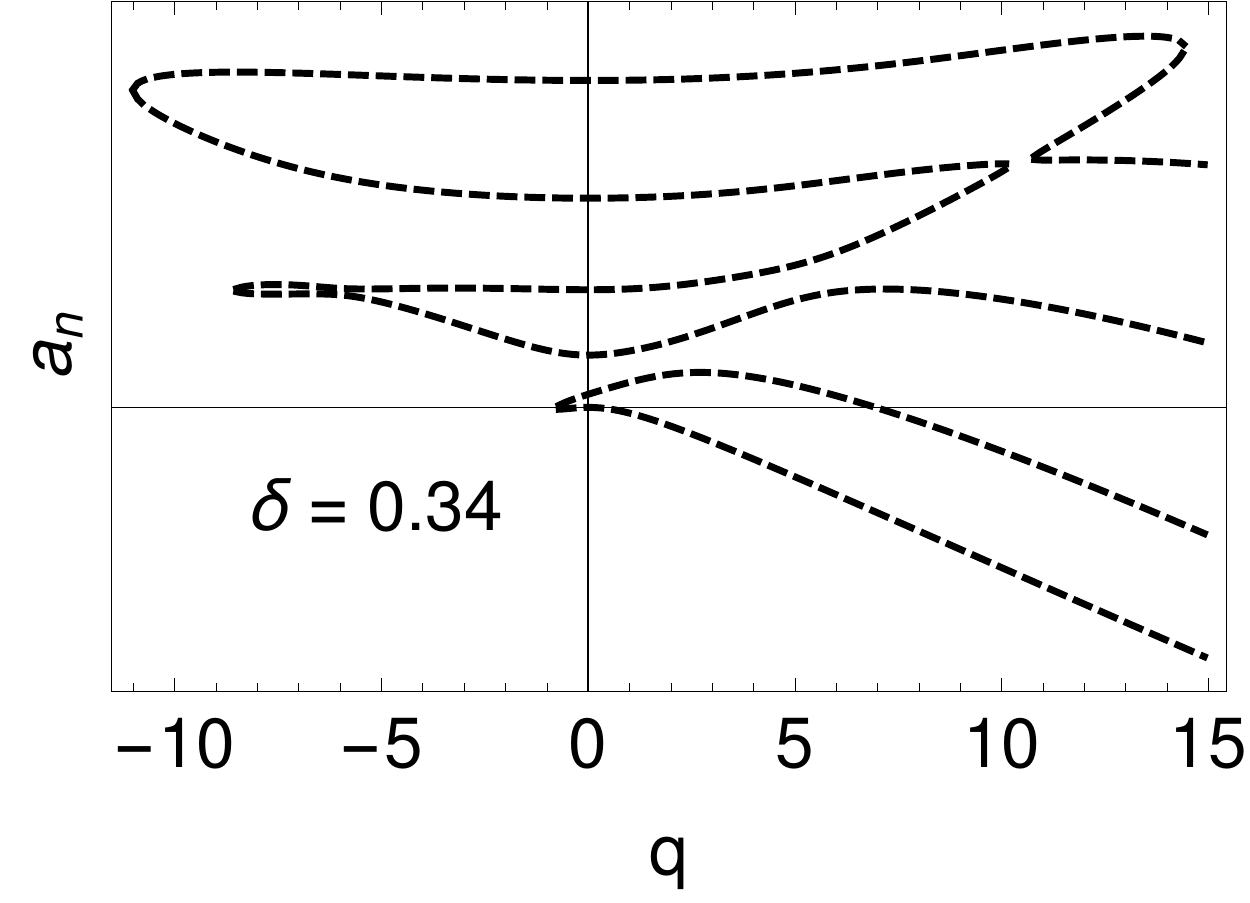}\\
		\includegraphics[width=0.45\linewidth]{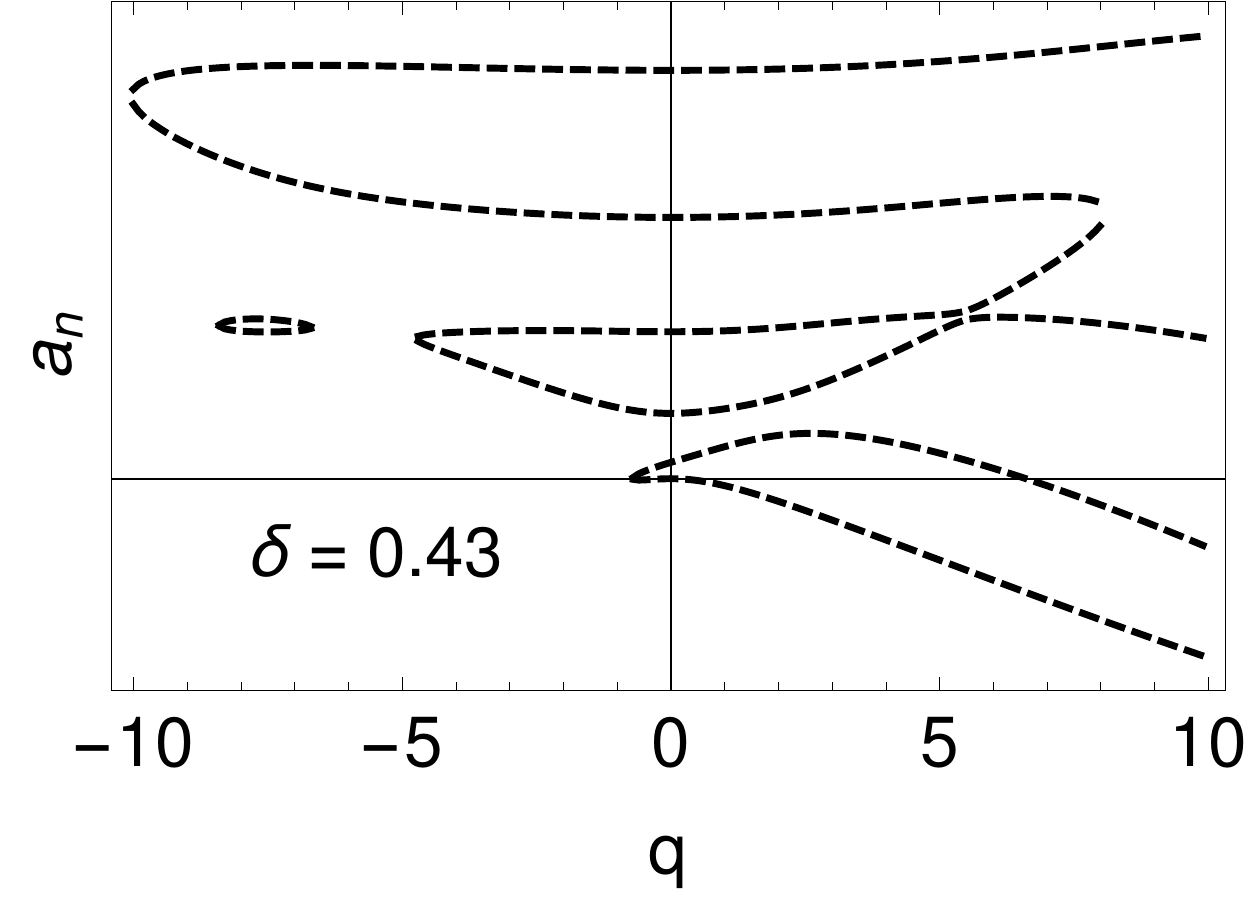}
		\includegraphics[width=0.45\linewidth]{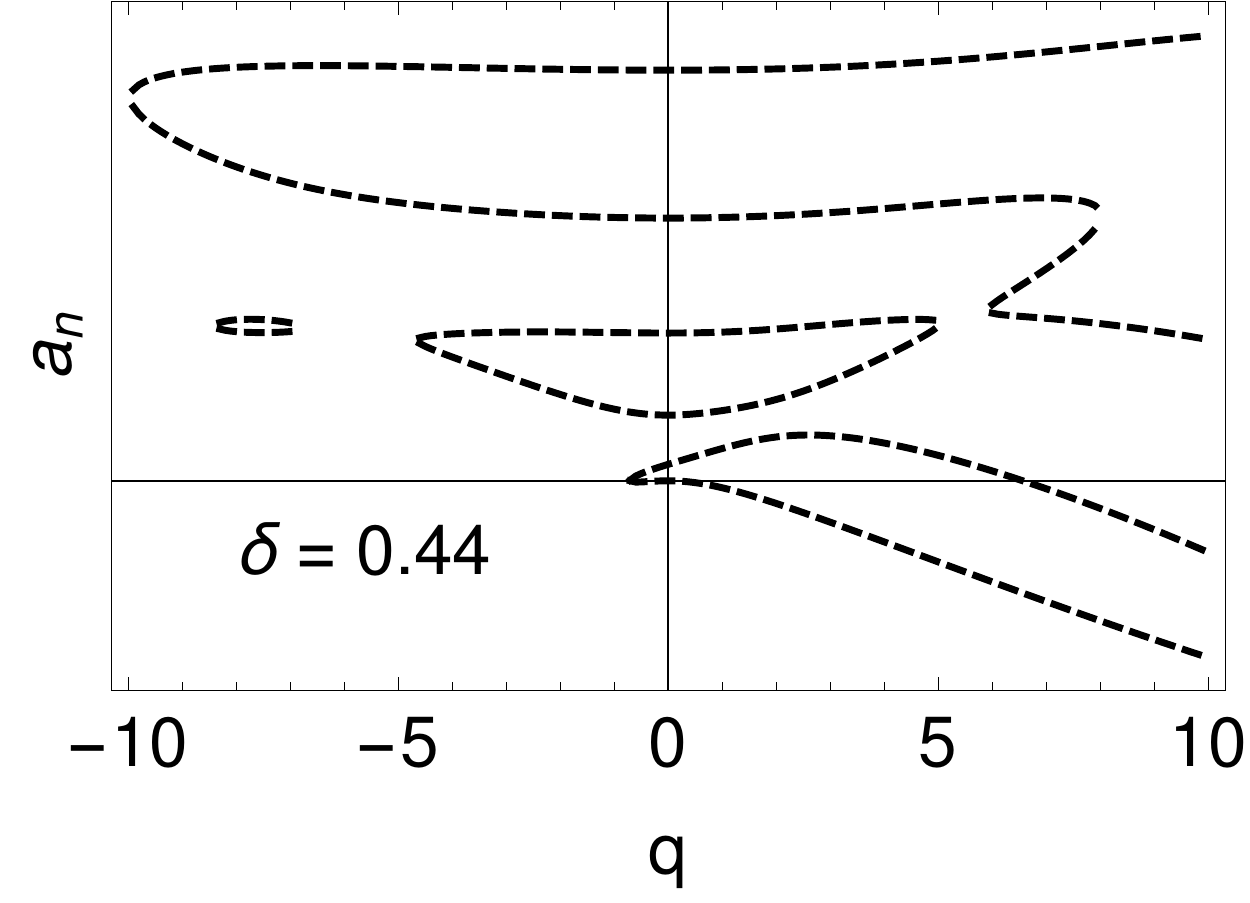}
		\caption{Energy levels $a_n$ ($u'(0)=u'(\pi)=0$) for $j=2$. (Upper left) For $\delta=0.33$, pairs of energy levels merge at the left plane. (Upper right) For $\delta=0.34$, four energy levels merge, changing the structure. (Bottom left)  For $\delta=0.43$, an ``isolated loop" appears at the left plane, two energy levels become degenerate at a point in the right plane. (Bottom right) For $\delta=0.44$, a new energy loop appears.}
		\label{fig:Levels_j2}
	\end{figure}
	
	In what follows we only show the behavior of the energy levels $a_n$ produced by the Neumann boundary conditions, Eq.~\eqref{eq:boundary_neumann}. The second set of energy levels produced by Dirichlet boundary conditions, Eq.~\eqref{eq:boundary_dirichlet}, has a similar behavior and do not add any useful information at this moment. We return to it at the end of the section to discuss the $\mathcal{PT}$-broken regime. 
	
	In Fig.~\ref{fig:Levels_j2} we can see that the scenario $j=2$ starts similarly to the $j=1$ case, the energy levels join for $q<0$ for small values of the deformation parameter $\delta$. However, the increase in the deformation parameter $\delta$ which controls the strength of the imaginary contribution to the Hamiltonian produces a richer structure for the spectrum. 
	
	With the increase of the deformation, we notice lines merging for $q>0$, as can be seen for $\delta=0.33$ (upper left in Fig.~\ref{fig:Levels_j2}), this is the energy loop behavior commented in the previous section and already reported in many works about $\mathcal{PT}$-symmetric models. However, we see a new behavior, where four originally independent energy levels merge as a single line for  $\delta=0.34$ (See the upper right image in Fig.~\ref{fig:Levels_j2}). Increasing a bit further the deformation we also see the emergence of new loops (still connected to the origin at $q=0$) and ``isolated loops" (that are disconnected from the origin $q=0$). Notice that around $\delta=0.43$ (bottom left in Fig.~\ref{fig:Levels_j2}) two energy levels approach each other and become degenerated. This degeneracy marks the formation of the closed loop. Also, notice that for $\delta=0.44$ (bottom right in Fig.~\ref{fig:Levels_j2}) the system is well defined (the first six energy levels are present) for small $q$, then at some point $q$ two of them disappear (this means a transition from a closed to an open system) and after a small increase in $q$ we recover all energy levels. 
	
	The behavior of the energy levels $a_n$ for $j=2$ tells us another different aspect in comparison with $j=1$. Recall that for $j=1$ the breaking of $\mathcal{PT}$ symmetry could be checked by looking only to the lowest energy level (the composition of $a_0,a_1$). This occurs because the lowest energy levels produces the smallest energy loop (smallest values of the lower and upper boundary $q$ given by the coalescence points). However, at $j=2$ this is not the case. We see clearly in Fig.~\ref{fig:Levels_j2} that at $\delta=0.44$ the first energy level is not even limited from the right and there is already a loop produced by joining the original 3rd and 4th energy levels. This indicates the need to change our methodology to check the $\mathcal{PT}$-breaking.
	
	\begin{figure}
		\centering
		\includegraphics[width=0.45\linewidth]{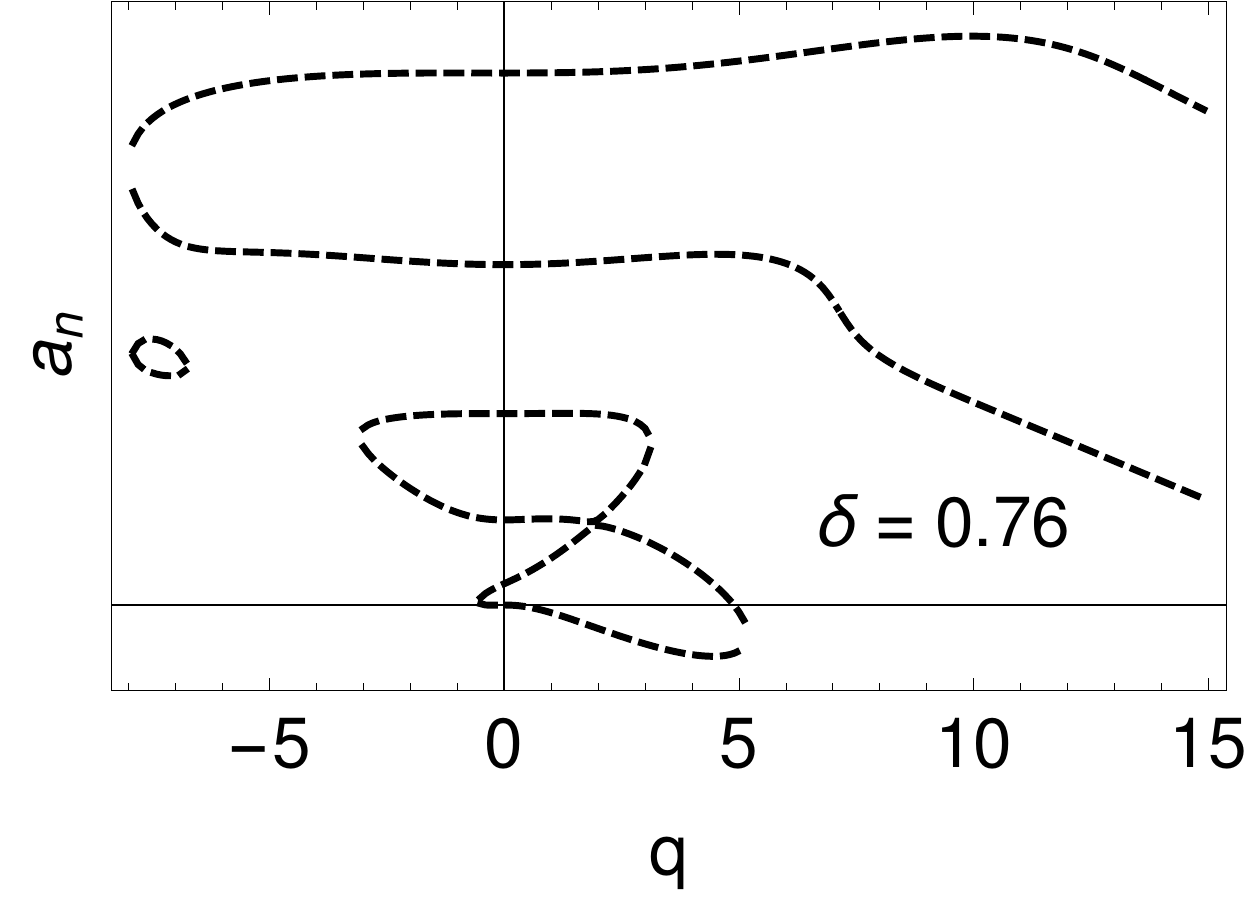}
		\includegraphics[width=0.45\linewidth]{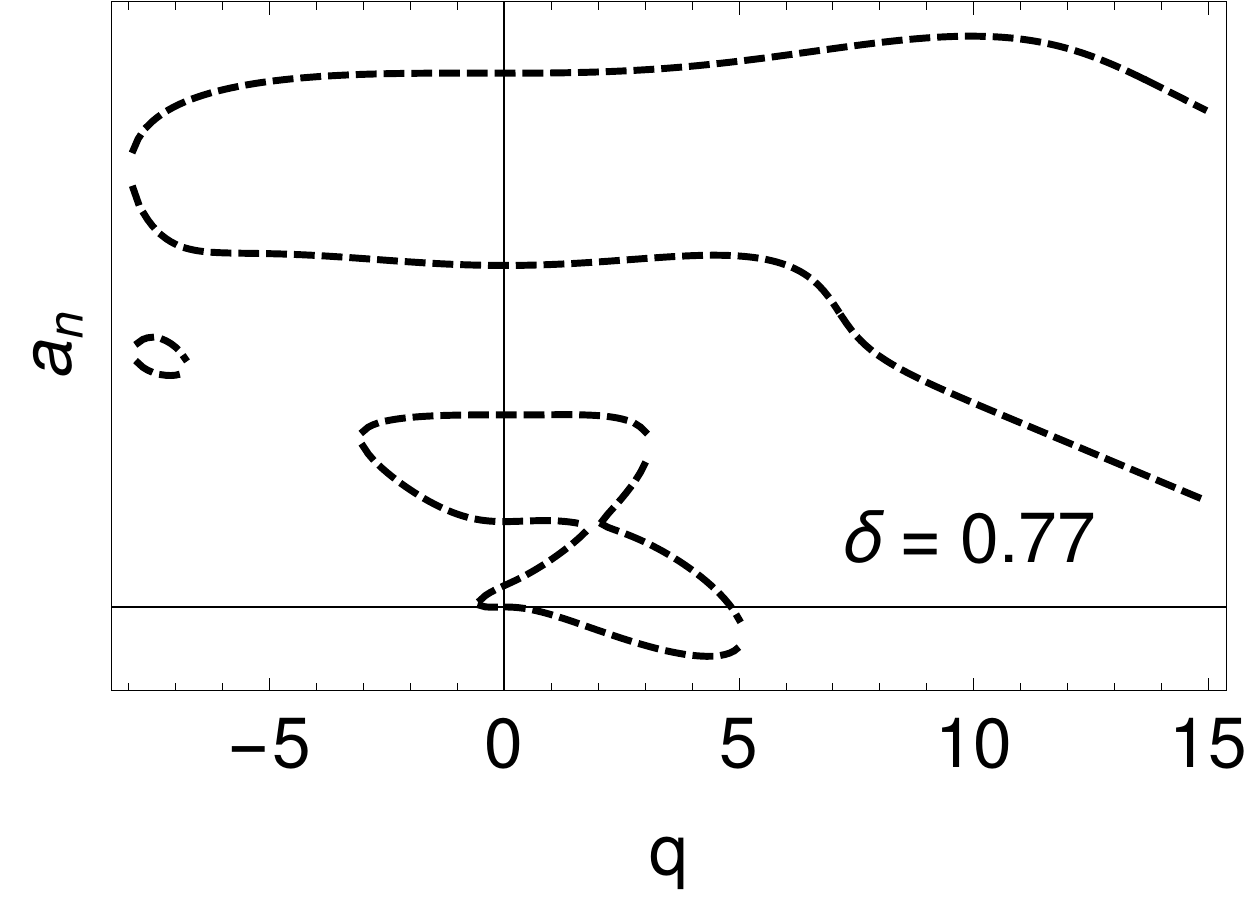}
		\caption{The $a_n$ energy levels for $j=2$. At $\delta=0.76$ (Left) the $\mathcal{PT}$-breaking occurs in the degeneracy joining the 3rd and 4th energy levels , at $\delta=0.77$ (Right) a jump occurs and the $\mathcal{PT}$-breaking occurs at coalescence point of the 2nd and 3rd energy levels.  }
		\label{fig:Levels_j2b}
	\end{figure}
	
\begin{figure}
		\centering
		\includegraphics[width=0.6\linewidth]{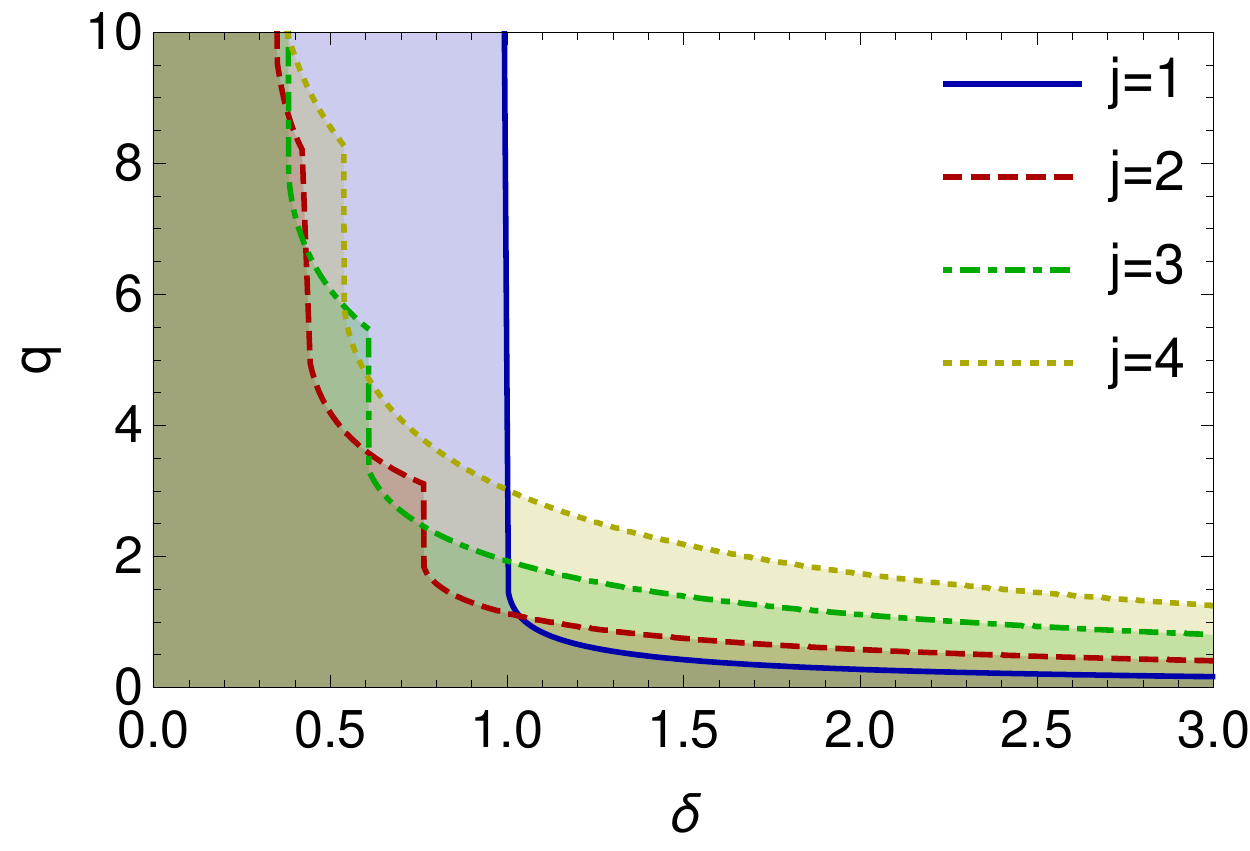}
		\includegraphics[width=0.6\linewidth]{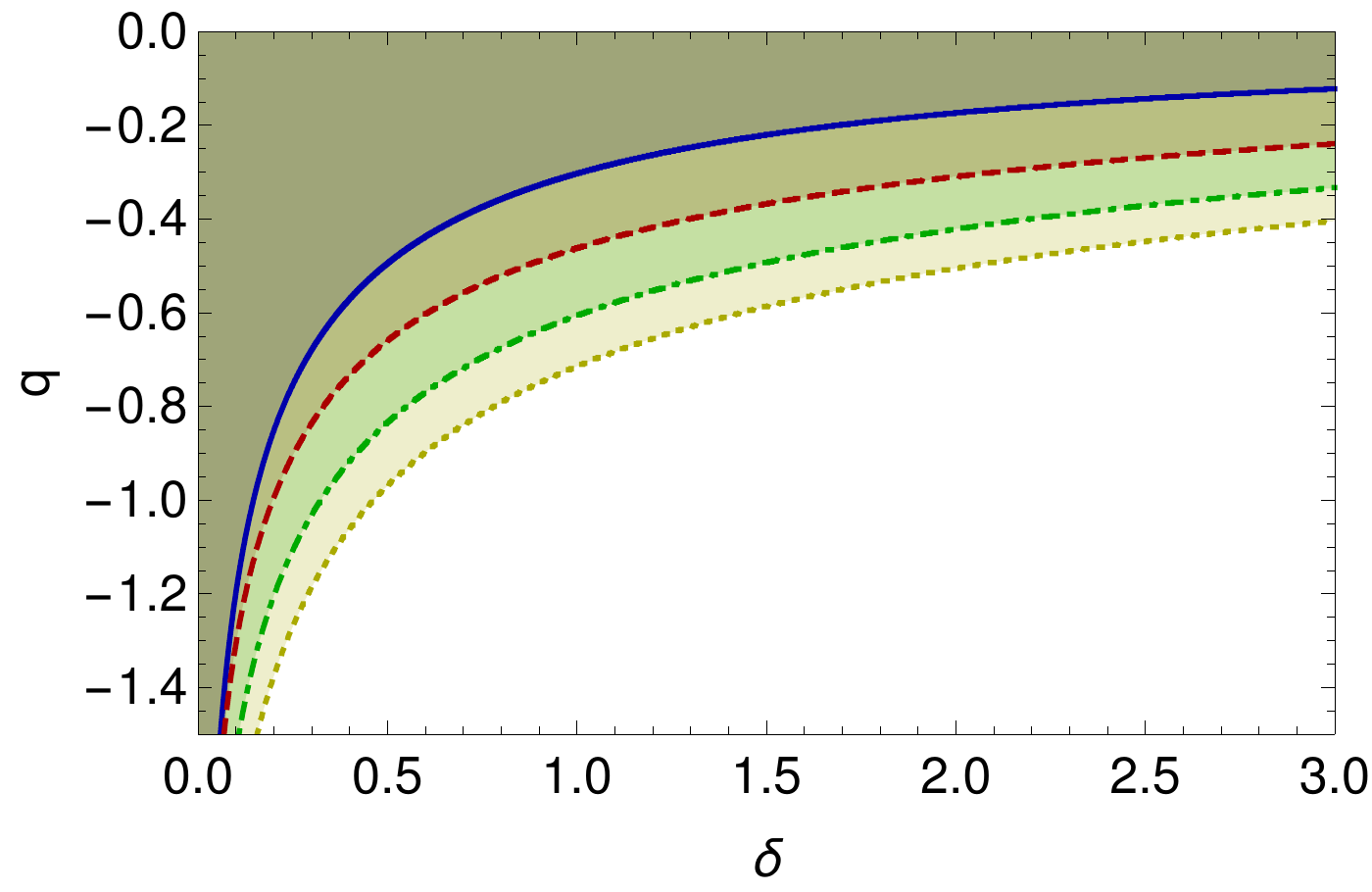}
		\caption{Exceptional lines at which a coalescence of energy levels $a_n$ occurs, for different values of the parameter $j$. (Upper) First quadrant of the $\delta \times q$, above each exceptional line we have a $\mathcal{PT}$-broken phase and at least one energy level becomes complex. (Bottom) Fourth quadrant of the $\delta \times q$; under each exceptional line $\mathcal{PT}$ symmetry is broken. We do not show the 2nd and 3rd quadrants because there is a reflection symmetry $\delta\rightarrow-\delta$ in the figure.}
		\label{fig:qcrit}
	\end{figure}

	The breaking of $\mathcal{PT}$ symmetry is controlled by the loop produced by the 3rd and 4th energy levels until this loop ``collides" with the loop produced by the 1st and 2nd energy levels. This occurs around $\delta=0.76$ (see Fig.~\ref{fig:Levels_j2b}, left) where a new degeneracy occurs. After this ``collision" both loops merge into a single energy loop with six coalescence points (see Fig.~\ref{fig:Levels_j2b}, right). This behavior will be responsible for a jump when we try later on to determine the dependence between $\delta$ and $q$. In the scenario exhibited in Fig.~\ref{fig:Levels_j2b} the jump occurs from the coalescence between the 3rd and 4th energy levels to the coalescence between the 2nd and 3rd energy levels. We can also see it in Fig.~\ref{fig:qcrit}.
	
\begin{figure}
		\centering
		\includegraphics[width=0.45\linewidth]{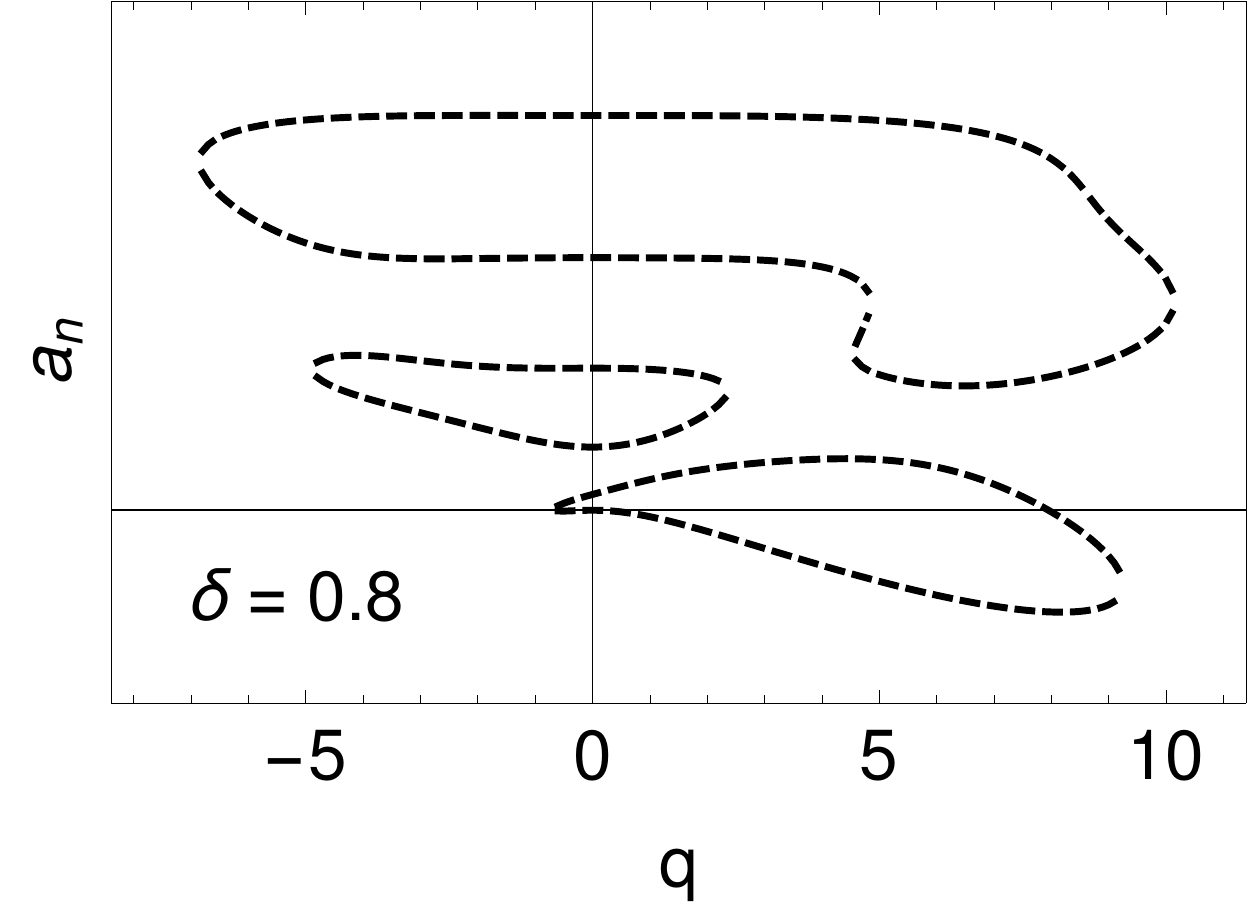}
		\includegraphics[width=0.45\linewidth]{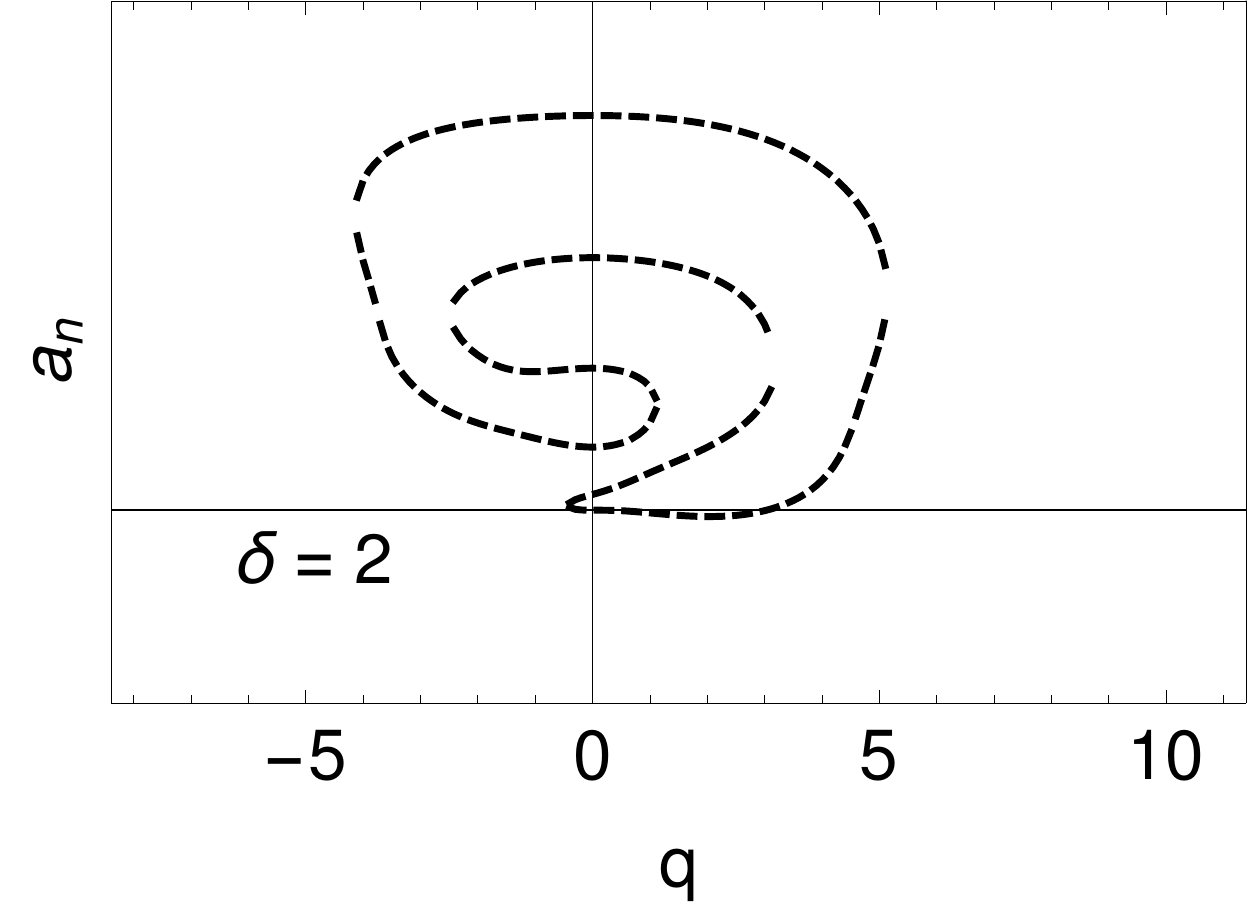}
		\caption{The $a_n$ energy levels for $j=3$. At $\delta=0.80$ (left) we see three closed loops produced by the first six energy levels. At $\delta=2$ (right) these energy levels merge into a single loop. Notice that in both figures the $\mathcal{PT}$-breaking is controlled by the coalescence of the 3rd and 4th energy levels.}
		\label{fig:Levels_j3}
	\end{figure}

	We remark that such discontinuity does not necessarily occur at any time at which independent energy loops merge. For example, at $j=3$, $\delta=0.8$ (Fig.~\ref{fig:Levels_j3}, left) the breaking point is signalled by the 3rd and 4th energy levels, and it is still the case even after the first six energy levels are merged into a single loop (Fig.~\ref{fig:Levels_j3}, right).
	
	Instead of looking at just one energy level, as we did for $j=1$, we see, based on the previous discussion, that we must check all the first energy levels and identify a $\mathcal{PT}$-breaking if anyone of them gets complex. For all scenarios the $a_n$ energy levels (Neumann b.c.) break first (small values of $q$) than the $b_n$ energy levels (Dirichlet b.c.). This is why we focus our attention first on the $a_n$ set. In Fig.~\ref{fig:qcrit} we see how the exceptional value of $q$ varies both with $\delta$ and $j$. Notice that for $j=2$ there is a discontinuous jump around $\delta\approx0.76$: this is exactly the jump we discuss on Fig.~\ref{fig:Levels_j2b}. These jumps also occur for different choices of $j$ as we increase the observed range of possible values of the parameter $q$.
	
	The curve $q(\delta,j)$ represents what we call an ``exceptional line", and it marks a transition between a phase of unbroken $\mathcal{PT}$ symmetry (under each curve, all energy levels are real) and a phase of broken $\mathcal{PT}$-symmetry (above each curve, at least one energy level is complex). This transition also means a transition between closed and open systems. It is important to notice that our diagram does not contemplate the recovery of the energy levels, although we know this behavior occurs as shown in Figs.~\ref{fig:Levels_j2}, \ref{fig:Levels_j2b} and \ref{fig:Levels_j3}. This was an active choice as we are mostly interested in the unbroken $\mathcal{PT}$ phase that is connected with the origin at $q=0$; in a more complete phase diagram, ``islands" of unbroken $\mathcal{PT}$ phases are expected.
	
	Regarding the ``stability" of the system, in the sense that we have a bigger parametric region $(q,\delta)$ where a $\mathcal{PT}$-symmetric phase connected to the origin occurs, we can see two main behaviors. For $q>0$ and small values of $\delta$, the increase of the factor $j$ means a loss of ``stability" as indicated in Fig.~\ref{fig:qcrit} by the diminished area under the exceptional line. Anyway, we cannot extract any clear hierarchy between different choices of the parameter $j$ due to the occurrence of the jumps. On the other hand, both for large values of the deformation $\delta$ and for negative values of the parameter $q$ we observe a clear hierarchy where the increase in the parameter $j$ means an increase in the parametric region with a $\mathcal{PT}$-symmetric phase.
	
	For large values of $\delta$ the exceptional lines exhibit a smooth behavior for all choices of the parameter $j$. 
	However, although the only free parameter is the value of $j$, it looks like the decaying behavior is not related to the value of $j$. If we make a power law fitting of the type $q(\delta,j) = A_j \delta^{-\alpha_j}$ for each exceptional line we get
	\begin{subequations}
		\begin{align}
		q(\delta,j=1) = 0.72\, \delta^{-1.37},\\ 
		q(\delta,j=2) = 1.07\, \delta^{-0.88},\\ 
		q(\delta,j=3) = 1.92\, \delta^{-0.79},\\ 
		q(\delta,j=4) = 3.03\, \delta^{-0.80}.
		\end{align}
		\label{eq:qdelta_fit}
	\end{subequations}
	
	\begin{figure}
		\centering
		\includegraphics[width=0.6\linewidth]{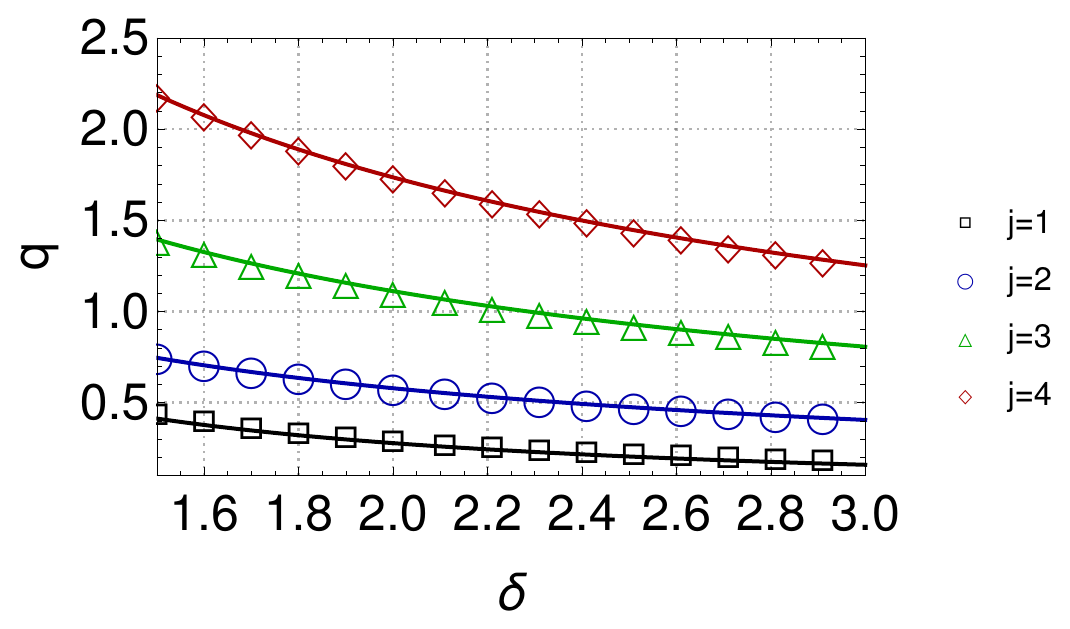}
		\caption{Power-law fit for the exception line $q(\delta,j)$ for different choices of $j$. The markers are numerical points (square for $j=1$, circle for $j=2$, triangle for $j=3$ and diamond for $j=4$) and the curve is the fitting $q(\delta,j) = A_j \delta^{-\alpha_j}$. See Eqs.~\eqref{eq:qdelta_fit} for the values of $A_j$ and $\alpha_j$.}
		\label{fig:qcritfit}
	\end{figure}
	\noindent However, as commented, the exponents of this power law do not seem to the authors to have any relationship with the value of the parameter $j$.
	
\begin{figure}
		\centering
		\includegraphics[width=0.6\linewidth]{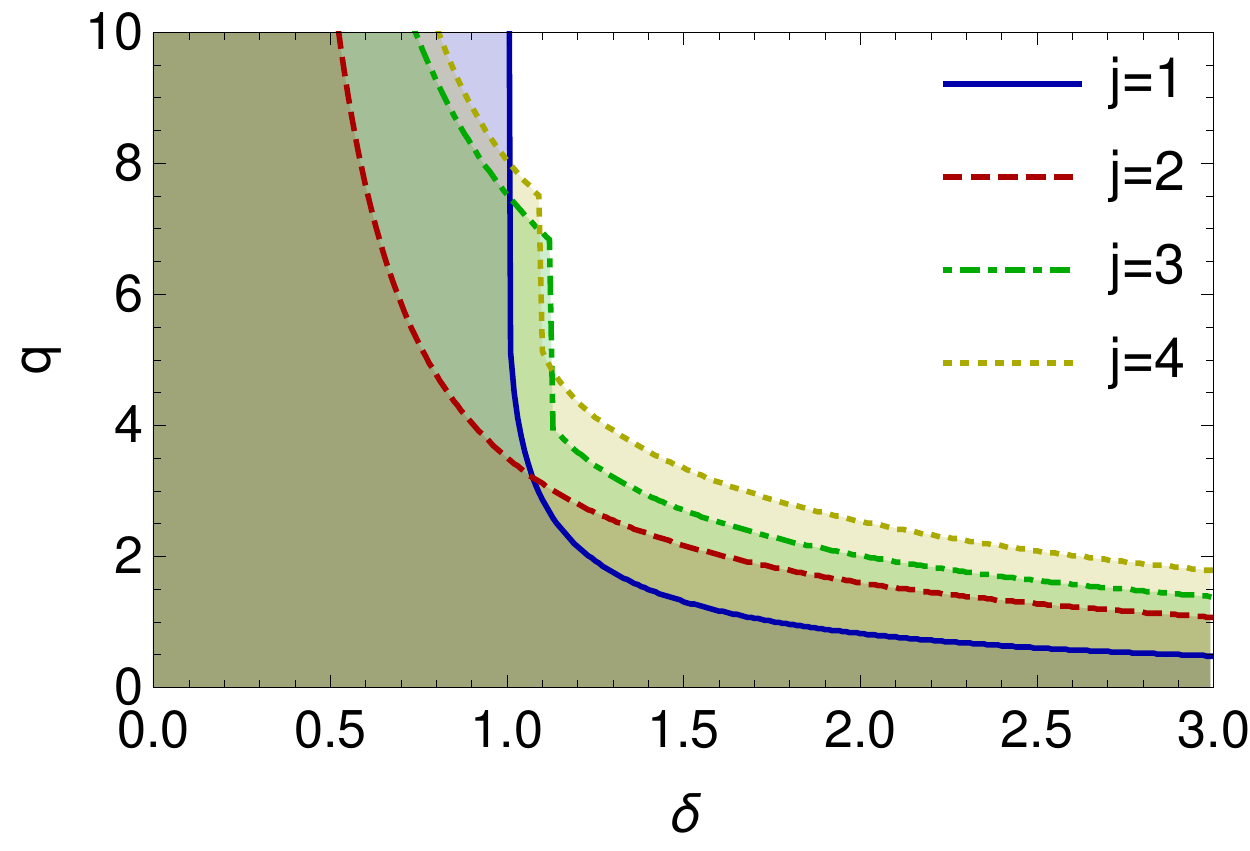}
		\includegraphics[width=0.6\linewidth]{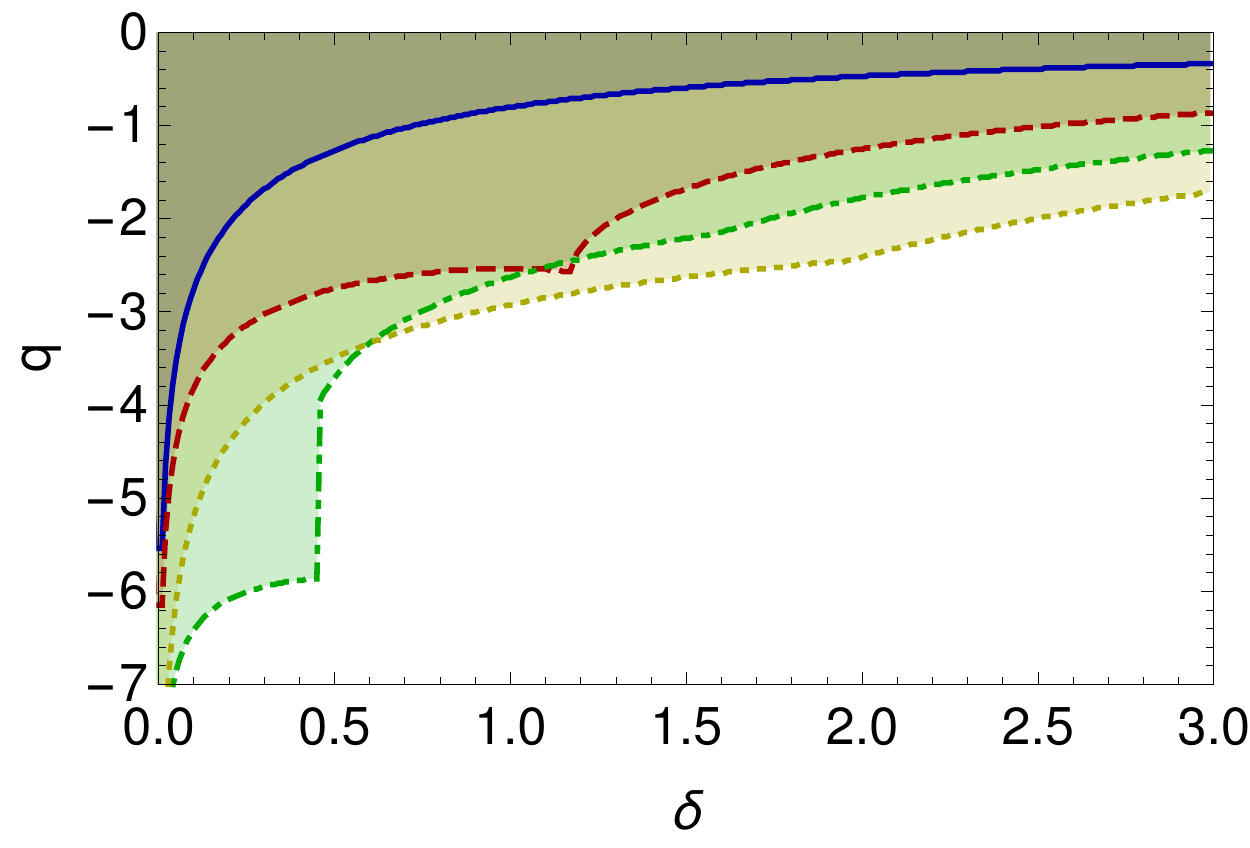}
		\caption{Exceptional lines at which a coalescence of energy levels $b_n$ occurs, for different values of the parameter $j$ in the $\delta \times q$ plane. In the filled region we have an unbroken-$\mathcal{PT}$ phase. Recall that there is a reflection symmetry $\delta\rightarrow-\delta$ in the figure.} 
		\label{fig:qcritbn}
	\end{figure}

	For completeness, let us now show, see Fig.~\ref{fig:qcritbn}, the exceptional lines for the second set of energy levels ($b_n$) produced by the Dirichlet boundary conditions, Eq.~\eqref{eq:boundary_dirichlet}. In this scenario, we do not see any discontinuous jump for $j=2$. Also, for all values of $j$ the region of unbroken $\mathcal{PT}$ is broader than the previous scenario (Fig.~\ref{fig:qcrit}, Neumann boundary condition). This means that for the deformed Mathieu equation the Dirichlet b.c. turns the system more stable than the Neumann boundary condition. Intuitively, we usually say that the Dirichlet b.c. means a \textit{closed} boundary condition while the Neumann b.c. is a \textit{open} boundary condition. In this sense, it is very natural that the Neumann b.c., as an open b.c., breaks $\mathcal{PT}$ first.
	
	\section{Conclusion}
	\label{sec:conclusion}
	
	We studied in this article a non-Hermitian but $\mathcal{PT}$-symmetric deformation of the Mathieu equation (which might be reproduced, for example, in optical experiments), that has as free parameters the deformation strength $\delta$ and a parameter $j$ related to the frequency of the imaginary contribution.
	We managed to show that the $\mathcal{PT}$-breaking point depends on the deformation parameter $\delta$ and we discussed a phase diagram where an exceptional line signals the transition from the unbroken $\mathcal{PT}$ phase to the broken $\mathcal{PT}$ phase. We also show that the change of the parameter $j$ turns the structure of the spectrum much richer than the simple $j=1$ scenario.
	
	Concerning the transition from a closed system (unbroken $\mathcal{PT}$ phase) to an open system (broken $\mathcal{PT}$) we see that:
	\begin{enumerate}
		\item The increase of $q$ reduces ``stability" and favors the transition. Recall that $q=0$ would make our model return to the classic harmonic oscillator.
		\item The increase of the deformation parameter $\delta$ also diminishes ``stability" and favors the transition. This is expected, as this is the parameter which controls the strength of the imaginary contribution to the Hamiltonian.
		\item The increase in the frequency, controlled by the parameter $j$ of the sine function makes ``stability" grow.
		\item Dirichlet boundary conditions produce a stabler set of energy levels than Neumann boundary conditions, which suggests that boundary conditions play a significant role.
	\end{enumerate}	
	
	These conclusions suggest possible experimental investigations to be performed for physical scenarios which can be modelled by a Mathieu equation. 
	Also, we discuss in our model, there is an impact of both the choice of boundary conditions and a parameter related to the frequencies. We indicate, from a formal perspective, the importance of investigating the different choices of $\mathcal{PT}$-symmetric extensions one can make for a selected model.

	\acknowledgments{
		E.C. thanks the Brazilian agency \textit{Conselho Nacional de Desenvolvimento Cient\'ifico e Tecnol\'ogico} (CNPq) for financial support through the PCID-B scholarship (\textit{Programa de Capacitação Institucional}). N.M.A. was financed by the \textit{Coordenação de Aperfeiçoamento de Pessoal de Nível Superior -- Brasil} (CAPES) -- Finance Code 001. 
	}
	
	\appendix
	\footnotesize
	\bibliography{0AllRefs}{}
\end{document}